# Power of Artificial Intelligence to Diagnose and Prevent Further COVID-19 Outbreak: A Short Communication

Muhammad Lawan Jibril[1,0000-0003-4175-424X,*,♯] and Usman Sani Sharif[2,0000-0002-7965-8678,*,♯]

[1]Department of Mathematics and Computer Science, Faculty of Science, Federal University of Kashere, P.M.B. 0182, Gombe- Nigeria

[2]Department of Biological Sciences, Faculty of Science, Federal University of Kashere, P.M.B. 0182, Gombe, Nigeria

*Correspondence: lawan.jibril@fukashere.edu.ng, ssu992@fukashere.edu.ng
♯These authors contribute equally

**Abstract**

Novel coronavirus-19 (2019-nCoV or COVID-19) is by far the most dangerous coronavirus ever identified for the third time in the three decades capable of infecting not only the animals but also the humans across the globe. Nearly 6000 deaths have been recorded due mainly to COVID-19 outbreak worldwide and more than 50% of these deaths appeared to have evolved from China where the virus was thought to originate. The endemicity of COVID-19 dramatically surpassed severe acute respiratory syndrome coronavirus (SARS-CoV) and Middle East respiratory syndrome coronavirus (MERS-CoV) that were so far discovered in 2003 and 2012 respectively. Thus, the World Health Organization (WHO) has declared the 2019-nCoV outbreak not only a public health emergency but also pandemic in nature. Currently, over 120 countries including Nigeria were reported to have more than 157,844 confirmed cases and 5,846 deaths due mainly to COVID-19 outbreak as of March 15, 2020, 10:55 GMT. Artificial Intelligence (AI) is widely used to aid in the prediction, detection, response, recovery of disease and making clinical diagnosis. In this study, we highlighted the power of AI in the containment and mitigation of the spread of COVID-19 outbreak in African countries such as Nigeria where human to human contact is apparently inevitable.

## INTRODUCTION

At the middle of December, 2019, novel Coronavirus (2019-nCoV or COVID-19) originated from Hubei Province, China called Wuhan [6, 12, 14, 35]. World Health Organization (WHO) has declared 2019-nCoV outbreak pandemic that negatively affected the world's economy [34, 2, 3]. The first case of international symposium on coronaviruses (CoVs) was held in 1980 and the major advances with regard to CoVs have emanated in 1983 due to application of molecular biology techniques including cloning and sequencing of the avian and murine genomes [32, 27, 28, 12, 15]. CoVs such as 2019-nCoV, severe acute respiratory syndrome coronavirus (SARS-CoV) and Middle East respiratory syndrome coronavirus (MERS-CoV) have been reported to cause a wide range of diseases in humans and animals [25, 31, 32, 38, 36, 12]. However, many researchers are of the opinion that, the virus originated from animals and spreads to human beings after which they infect each other [23, 24, 4]. CoV is likely the third time in three decade to jump from infecting animals to human beings [16]. SARS and MERS-CoVs are coronaviruses that were discovered in 2003 and 2012 respectively and are responsible for acute respiratory syndrome in humans in the Middle East, Europe, North Africa, and the United States of America [35, 36, 12]. However, the severity of 2019-nCoV appears to dramatically exceed SARS and MERS-CoVs [12, 16, 38, 36, 18]. According to Word Meter, today 15$^{th}$ March, 2020, 10:55 GMT, there are 157,844 confirmed cases and 5,846 deaths from the coronavirus COVID-19 outbreak [30] and the rate at which confirmed and dead cases of COVID-19 increases, is indeed exponential. Although the outbreak of 2019-nCoV in China appears now to be slowed down and delayed, cases outside China – most notably in Italy, Iran, Spain, France and United Kingdom are increasing at an alarming rate as dead cases due to COVID-19 surpassed 1800, 720, 280, 120 and 40 respectively [34, 35, 27]. Currently, over 120 countries have been reported to have 145,015 confirmed cases of 2019-nCoV with approximately 5,408 deaths [30]. Of more than 145,015 confirmed cases, lowest laboratory confirmed cases emanated from African countries including Nigeria, South Africa, Algeria, Rwanda, Namibia, Tunisia, Egypt, Senegal, Kenya, Congo, Sudan and Ethiopia – in most cases as a result of importation by foreigners visiting or working in the aforementioned countries [34, 35, 2, 13].

The first case of the disease in sub-Africa has been confirmed in Lagos State – Nigeria on 25$^{th}$ February, 2020 where Italian citizen who works in the country appears to be tested positive for COVID-19 [2, 35]. Nigeria has so far recorded 3 confirmed cases of COVID-19 and the second confirmed case which was as a result of contact of index case has now been tested negative and the first confirmed case has been by far clinically stable and the third case has been reported today, 17$^{th}$ March, 2020 in Lagos State, Nigeria and the patient is independent of the index case [2, 34]. WHO warned that African's fragile health systems meant the threat posed by the disease was considerable [23–25]. According to WHO, outbreak of coronavirus has reached a decisive point and has pandemic potential which threatens the lives of people [34, 35, 2]. Therefore, it is clear that right now the world needs speedy and quick solution to diagnose and tackle further spread of COVID-19 especially in developing countries including Nigeria. The power of artificial intelligence (AI) can be leveraged to limit the further outbreak of COVID-19 which entails the incorporation of AI for the containment and mitigation with a view to preventing the pandemicity of trending outbreak especially in African countries where human to human contact is apparently inevitable.

Although the major symptoms of COVID-19 are fever and a cough as well as short of breath which in many instances appeared to be similar to that of flu [6, 7, 24], the AI Based System can pinpoint the COVID-19 from computed tomography images in about 15 seconds with 90% accuracy compared to 15 minutes by human physicians [17]. Similarly, the applications of AI to diagnose and prevent further outbreak of COVID-19 especially to developing nations including Nigeria are of epidemiological importance for the containment and mitigation of COVID-19 outbreak especially in African countries where human to human contact is apparently inevitable. In this study, the applications of AI to diagnose and prevent further outbreak of COVID-19 have been highlighted, reviewed and discussed and some further observations to developing nations such as Nigeria were recommended on how to harness the power of AI to diagnose and prevent further 2019-nCoV outbreak in their respective countries.

**POWER OF AI**

The term artificial intelligence (AI), sometime called machine intelligence, refers to simulation or imitation of human intelligence in machine to think or act like human [19, 20]. It applies to any machine or system that exhibits traits associated with a human such as problem solving or learning as the case may be [14, 19]. Health sector or industry in most of the developed nations has always been a leader in innovation with technology [14]. However, the rapidly mutating of diseases such as 2019-nCoV, SARS-CoV and MERS-CoV makes it difficult to stay ahead of the curve, but with the aid of AI, the health sector or industry continues to advance thereby creating new predictive, diagnostic and preventive systems for the treatment of these new trending diseases especially COVID-19 with the aim of helping people live healthier and longer [20, 21]. CoV was predicted by AI long before the world really knows much about it [4, 5, 20, 21]. Intelligence system that sifts the data about people was developed by AI Startup Company called BlueDot, determines by far the chances of disease occurrence [4, 5]. The team of BlueDot confirmed that, the information its systems has relayed and informed its clients predicted 2019-nCoV pandemic long before Chinese and WHO made official announcement [4, 5]. The system predicted 2019-nCoV pandemic and turned out to be true. The system alerted the outbreak of the 2019-nCoV in December, 2019 and later became true as the outbreak appeared to be the mainstream on 2nd February, 2020 [9, 4, 5]. This shows and attests the powerful nature of AI technology. Therefore, the power of the AI cannot be underestimated, had it been the prediction of the system had been acknowledged many people around the world would not have been the victims of COVID-19.

Although the major symptoms of COVID-19 are fever and a cough as well as short of breath, flu shares many similar symptoms thereby making 2019-nCoV very difficult to be diagnosed without a test [6, 7, 24]. Similarly, the whole process of testing 2019-nCoV takes 5 – 7 hours [15, 7]. However, AI Based Smart image reading system to diagnose COVID-19 has been developed by China based company [1, 4]. The system reads the CT images in seconds rather than minutes [1, 4, 8]. The system is already being used to scan more than 5000 patients [7]. The AI Based System can pinpoint the COVID-19 from computed tomography images in about 15 seconds with 90% accuracy compared to 15 minutes by human physicians [17]. The analysis engine of the AI Based Smart Image Reading System measures changes in lesion, tracks development of COVID-19 and thus evaluates treatment [17, 4]. The system is also a cloud based which can be accessed by radiologists via cloud [17]. AI Based Standing X-ray Machine is currently building by Nanox that will supply tomographic images of the lungs [26]. The machine is to be installed in public places such as train stations, airports, seaports, land borders or anywhere else where large groups of people rub shoulders [21, 26]. The system attracted investments funds to capitalize on AI's potential for thwarting COVID-19 epidemics as such Foxconn invested $26 million to supply 1,000 systems to medical imaging services across New Zealand, Australia and Norway [26, 21, 22]. Therefore, this is one of the greatest testimonies to how the future of the epidemic prevention of 2091-CoV lies on the power of AI. AI Based System is currently being developed by In-Silico Medicine, Startup Company [8, 3]. The system will be informing doctors about the molecules that are capable of fighting COVID-19 [4]. AI Based System is fast and accurate having recently analyzed molecules and provided feedback about the molecules suited to counter the virus [3, 4]. The company is developing database of molecules that medical researchers can use in their projects and more so to use the database towards combating the further outbreak of the 2019-nCoV [4]. John Brownstein who is the chief officer at Harvard Medical School used AI techniques called machine leaning to comb through news reports, social media posts and data from official public heath channels and information sourced from doctors for warning signs of 2019-nCoV taking hold in countries outside China [1]. The AI Based System can search online information about 2019-nCoV and understand locations of the outbreak of the disease. The system therefore facilitates the discovery of the locations of outbreak so also promotes increased enlightenment and awareness of the potential solutions [33, 1].

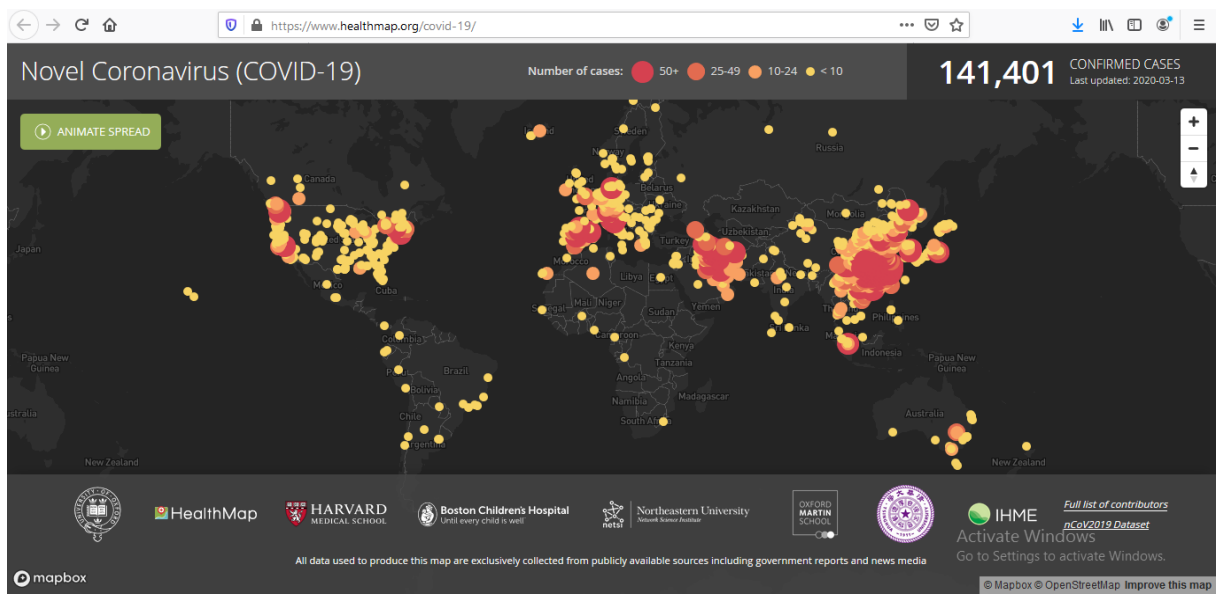

Healthmap's data visualization of 2019-nCoV's spread across the word as 14[th] March, 2020
Source: www.healthmap.org/covid-19/

Danish Company supplied Disinfecting AI Based robots manufactured for China and the robots rove around health facilities spreading ultraviolent light to disinfect rooms contaminated with COVID-19 [4]. Use of the robots limits the exposure of medical and other personnel to 2019-nCoV [11]. Researchers from Wuhan EndoAngel, a Medical Technology Company, Renmin Hospital of Wuhan University and China Geosciences used deep learning AI concept to detect 2019-nCoV [4]. The accuracy of the model in their study is 95%, where the model trained with CT scan of 51 patients with laboratory confirmed COVID-19 pneumonia and more than 45,000 anonymized CT scan images [8]. The model showed a performance comparable to radiologist and also improved efficiency of radiologists in clinical practice and the model can decrease the confirmation time from CT scans by 65% [29, 11]. AI robots were manufactured by tech companies in China to remind people to wear masks for COVID-19 prevention [11]. Predictive analytics from data is indeed changing the approach of the outbreak management by updates which are often followed by additional updates [29, 10].

The researchers and scientists are leveraging AI to predict hotspots where new disease could emerge [10, 27]. WHO says Europe is now the epicenter of COVID-19 [34, 35, 2] and this conclusion is likely to be predicted using AI. This clearly shows that AI is being leveraged to prevent the further occurrences of COVID-19 [26]. AI can be used to integrate about known virus, human demographic, animal population and socio-cultural practice around the globe to predict further outbreak of 2019-nCoV or even other unknown viruses [29]. Therefore, government especially African governments and that of other developing countries should use such kind of data to be proactive or to take further steps and action towards preventing such kind of outbreak or to prepare for such kind of calamity [34]. AI is speed and scale in terms of preventing pandemic of COVID-19 [35]. It detects changes in condition instantly which makes AI a greater source of power when trying to prevent COVID-19 pandemicity [26]. The AI system may scan 60 people per day on average, therefore, with 1000 systems, 60,000 people can be scanned on daily basis. Therefore, no human resources available today can support this volume of diagnosis with efficiency and speed [7, 14]. The faster and quicker an outbreak of the disease is detected, the sooner action can be taken to stop its further spread and treatment to the infected ones effectively [14]. Prompt response to outbreak of pandemicity of 2019-nCoV after it has been identified, can be seen and testified in some of the AI Based Application so far developed which limit the impact of the disease [7]. AI integrates wide range of travelling, population and disease data to predict how and where it might spread for proper and prompt response as likely predicted by WHO that Europe is now epicenter of COVID-19 [29, 2]. AI Deep learning is used by radiologists to build a model that learn from experience with dataset so as to make treatment decision based clinical and medical imaging [37]. Lastly, once the outbreak come to an end, WHO and governments around the world should make a turnaround strategy or decision on how to further limit or prevent outbreak of COVID-19 or other viruses yet to be identified. AI Machine learning algorithm can be used to simulate different outcomes so as to validate and test the decisions, policies or response plans.

**CONCLUSION**

Taken together, AI is extremely potent to the needful in every aspect of human life, as such, harnessing its power to diagnose and prevent further outbreak of COVID-19 around the world is of utmost importance not only for the containment and mitigation of COVID-19 but also for cushioning the effects of economic meltdown of some countries whose national economic strength depends largely on crude oil like Nigeria. AI can be leveraged to limit the further outbreak of 2019-nCoV in four ways namely: prediction, detection, response and recovery; thereby containing COVID-19 in both animate and inanimate nature. Thus, AI can be perceived as a visual guide for the containment and mitigation of the pandemicity of COVID-19 outbreak owing to the way and manner it secures our future.

**Funding**

No funding sources

**Conflict of interest**

Authors have declared that no conflict of interest exists.